\documentstyle[aps,amsfonts,amssymb,prl,psfig]{revtex}

\begin{document}
\twocolumn
\noindent
{\bf Comment on ``Non--equilibrium Electron Distribution in Presence of
Kondo Impurities''\\ (cond-mat/0102150.v2)}

In a recent paper \cite{goeppert.01} G\"oppert and Grabert (GG) 
investigate the effect of
inelastic scattering by Kondo impurities on the non-equilibrium distribution
function $f(\varepsilon, U)$ in resistive nanowires. 
The experimentally observed \cite{pothier.97,pierre.00} 
scaling of $f(\varepsilon, U)$ w.r.t.\ the bias 
voltage $U$ points at a $1/\omega^2$ behavior of the 
collision kernel $K(\omega,\varepsilon,\varepsilon ')$ 
\cite{pothier.97}, with $\omega$ the energy exchanged in the 
scattering process and $\varepsilon$, $\varepsilon '$ the particle energies.
This is explained by spin scattering 
in 4th order perturbation theory (PT) in the Kondo coupling $J$
\cite{glazman.01,kroha.00,goeppert.01}. Going beyond bare PT,  
(1) the infrared divergence of $K(\omega,\varepsilon,\varepsilon ')$
is cut off by a non-equilibrium decoherence rate $1/\tau _s$  and
(2) there are retardation effects arising from vertex renormalizations
\cite{glazman.01,kroha.01}. 

GG factorize the corresponding convolutions in the electron-electron
t-matrix by setting 
$\omega =0$
everywhere except in the leading infrared divergent factor $1/\omega ^2$.
This is valid only when $1/\tau _s  \ll U$ 
(i.e. when $U \gg T_K$), 
since then the $\omega$-integral over the range $|\omega | \lesssim U$ 
is dominated by the $\omega =0$ contribution. 
However, in the experiments, where $T_K$ is estimated \cite{pierre.00,kroha.01} 
such that  $U \approx 10 \dots 50 T_K$, one has 
$1/\tau _s \simeq (2\pi/3)\ S(S+1)  
[{\rm ln} (U/T_K)]^{-2}U \approx 0.3 \dots 0.1 U$, 
i.e.\ the factorization is not valid here. Neglecting $1/\tau _s$ 
throughout their calculation and factorizing the collision kernel,  
GG overestimate $K(\omega , \varepsilon, \varepsilon ')$, especially
for energy exchange $\omega \lesssim 1/\tau _s$. 
This provides an explanation why the Kondo impurity concentration
$c_{imp}$ used by GG in order to describe the experiments
is lower than $c_{imp}$ determined in Ref.\ \cite{kroha.01}, where 
$1/\tau_s$ is included and no factorizations are applied.

GG further emphasize that for $T\ll T_K$ their results do not depend on $T_K$, 
in contrast to Ref.\ \cite{kroha.01}, where the renormalized coupling 
depends on $U/\text{max}[T,T_K]$, 
and max$[T,T_K]$ sets the scale where deviations 
from scaling of $f(\varepsilon ,U)$ occur towards lower bias. 
The latter is obvious on physical grounds,
since for $U<T_K$ the Kondo impurity crosses over to a potential
scatterer, where $K\propto 1/\omega ^2$  must terminate.
Hence, the $T_K$--independence of GG's results is incorrect
and appears to be an artefact
of their factorization scheme and neglection of $1/\tau_s$.

Finally, GG claim that in the slave boson (SB) method 
employed in Refs.\ \cite{kroha.00,kroha.01} algebraic behavior 
of $K(\omega , \varepsilon, \varepsilon ')$ can only 
be obtained in infinite order PT.
We now show that, in contrast, the SB technique 
reproduces the $1/\omega ^2$ behavior of \cite{glazman.01,goeppert.01} 
in 4th order PT.
As pointed out in \cite{kroha.01}, 
\begin{figure}
\centerline{\psfig{figure=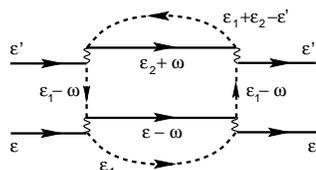,width=0.48\linewidth}} 
\caption
{Contribution to the electron-electron vertex of $O(J^4)$, included in NCA
(notation as in Ref.\ [6]).}
\label{fig1}
\end{figure}
\noindent
the in- and out-scattering 
single-particle t-matrices $t^<$, $t^>$
of $O(J^4)$ are included in NCA and are given by the diagram Fig.\
\ref{fig1} and the one with $\varepsilon \leftrightarrow
\varepsilon '$, $\omega\to -\omega$, where the bare pseudo-fermion and boson
propagators, $G_f^{r\ (0)}(\varepsilon ) =1/(\varepsilon +i0)$ and
$V^2 G_b^{r\ (0)}(\varepsilon\approx 0) = J$, are inserted and the lines 
carrying $\varepsilon '$ are connected \cite{note}. Hence
\begin{eqnarray}
t^> (\varepsilon ) &=& - \frac{8\pi i  }{\rho} c_{imp} (\rho J)^4
\int d\varepsilon ' \int d\omega \frac{1}{\omega ^2} 
\nonumber\\
&\times& [1-f(\varepsilon ' +\omega )] [1-f(\varepsilon -\omega )] 
f(\varepsilon ')   \ ,
\end{eqnarray}
and similar for $t^<$, with $f \leftrightarrow -(1-f)$.
Inserting this into the collision integral Eq.\ (2) of  
\cite{kroha.01}, one recovers the $O(J^4)$ expressions of 
\cite{glazman.01,goeppert.01} with the collision kernel \cite{note2}
\begin{eqnarray}
K(\omega,\varepsilon,\varepsilon ') = 
\frac{8\pi  (\rho J)^4}{\rho} \ \frac{c_{imp}} {\omega^2} \ .
\label{kernel}
\end{eqnarray}  

A correct treatment must incorporate decoherence and retardation.
Both are included in NCA through 
selfconsistency \cite{kroha.01} (see Fig.~\ref{fig1}), 
the former via the infrared cutoff $1/\tau _s$ of the pseudo-fermion 
propagators $G_f(\varepsilon _1 -\omega )$, the latter through 
the boson propagators $G_b(\varepsilon _1 -\varepsilon ' -\omega )$ 
which acquire frequency dependence by renormalization. 
Approximate scaling of $f(\varepsilon, U)$ at large $U$ does persist 
{\it despite} the cutoff $1/\tau _s$, because for $U> T_K$
$1/\tau _s$ itself scales with $U$, up to log corrections, and because 
retardation effects are resummed to yield algebraic
behavior of $t^{\gtrless} (\omega)$ \cite{kroha.01}. 
Deviations from scaling towards smaller bias, 
which are also observed experimentally \cite{pothier.97}
due to finite $T$ or $T_K$, are caused by $\ln(U/\text{max}[T,T_K])$ 
corrections both to
$1/\tau _s$ and to the algebraic behavior of $t^{\gtrless} (\omega)$.

We  thank the authors of Refs.\ \cite{pothier.97,pierre.00}, 
B.~Al'tshuler, J.~v.~Delft, 
L.~I.~Glazman, A.~Rosch, and P.~W\"olfle for valuable discussions.
This work is supported by DFG through SFB195, by grants
OTKA T024005, T029813, T034243 and by the A. v. Humboldt foundation.\\

\noindent
Johann Kroha$^1$  and Alfred Zawadowski$^2$ \\
{\small
${}^1$Institut f\"ur Theorie der Kondensierten Materie, \\
\hphantom{${}^1$}Universit\"at Karlsruhe, 76128 Karlsruhe, Germany\\ 
${}^2$Department of Physics and Hungarian Academy of Sciences,\\ 
\hphantom{${}^1$}Technical University of Budapest,\\
\hphantom{${}^1$}P.O.B. 6980, H--1525 Budapest, Hungary\\

\noindent
Dated: 2 May, 2001
}


\begin{thebibliography}{}
\bibitem{note}
All other diagrams of order $\leq 4$ are less divergent.
\bibitem{note2}
The prefactor differs from Refs.\ \cite{glazman.01,goeppert.01} 
because the Anderson model allows inelastic
scattering without spin flip.
\end{thebibliography}
\end{document}